\documentclass{iopconfser}

\usepackage{graphicx}
\usepackage{dcolumn}
\usepackage{color}
\usepackage{bm}
\usepackage{amsmath,amssymb,graphicx}
\usepackage{epsfig}
\usepackage{enumerate}

\begin{document}

\title{First integrals of some two-dimensional integrable Hamiltonian systems}

\author{Aritra Ghosh and Akash Sinha}

\affil{School of Basic Sciences, Indian Institute of Technology Bhubaneswar, Jatni, Khurda, Odisha 752050, India}

\author{Bijan Bagchi}

\affil{Department of Mathematics, Brainware University, Barasat, Kolkata, West Bengal 700125, India}

\email{ag34@iitbbs.ac.in}

\begin{abstract}
In this paper, we discuss some results on integrable Hamiltonian systems with two degrees of freedom. We revisit the much-studied problem of the two-dimensional harmonic oscillator and discuss its (super)integrability in the light of a canonical transformation which can map the anisotropic oscillator to a corresponding isotropic one. Following this, we discuss the computation of first integrals for integrable two-dimensional systems using the framework of the Jacobi last multiplier. Using the latter, we describe some novel physical examples, namely, the classical Landau problem with a scalar-potential-induced hyperbolic mode, the two-dimensional Kepler problem, and a problem involving a linear curl force. 
\end{abstract}

\section{Introduction}\label{introsec}
In the context of ordinary differential equations, the notion of integrability primarily rests on the existence of first integrals, also called constants of motion or conserved quantities; the existence of a sufficiently-many independent conserved quantities ensures integrability. Hamiltonian systems admit the notion of Liouville integrability: on a phase space of dimension \(2n\) (\(n \in \mathbb{N}\)), integrability requires \(n\) globally-defined and functionally-independent conserved quantities in involution, i.e., with mutually-commuting Poisson brackets \cite{fasso,evans,marquette}. One could (and often does) have more than \(n\) conserved quantities; such systems are called superintegrable, e.g., the Kepler problem. Since the phase space is of dimension \(2n\), the system may admit a maximum of \(2n-1\) functionally-independent conserved quantities; such systems are termed as maximally superintegrable. The purpose of this paper is to discuss certain computations of first integrals in the case of two-dimensional Hamiltonian systems, expanding upon previous developments by the present authors as reported in \cite{ANISO,2d16} (see also, \cite{2da,2db,2dc,ANISOprev}). With this (very) brief introduction, let us begin with the case of the two-dimensional harmonic oscillator.

\section{Two-dimensional harmonic oscillator}\label{sec2}
The two-dimensional harmonic oscillator is described by a separable Hamiltonian which takes the following form \cite{ANISO}:
\begin{equation}\label{Hisopq}
H =\sum_{j = 1}^2 H_j, \quad \quad H_j =  \frac{\omega_j}{2} \big(p_j^2 + q_j^2 \big),
\end{equation} where \(\omega_1,\omega_2 > 0\) are the frequencies on the \(q_1-p_1\) and \(q_2-p_2\) planes, respectively. Let us perform another set of canonical transformations that read
\begin{equation}\label{XP}
X_j = \frac{q_j - i p_j}{\sqrt{2}}, \hspace{5mm} P_j = \frac{p_j - i q_j}{\sqrt{2}},
\end{equation} and then the Hamiltonian gets expressed as \(H = i  \sum_{j=1}^2 \omega_j P_j X_j\), and which may be rewritten as 
\begin{equation}\label{HPX}
H = i \omega_0  \sum_{j=1}^2 \Omega_j P_j X_j, 
\end{equation} where
\begin{equation}
\Omega_j = \frac{\omega_j}{\omega_0} \quad \implies \quad  \frac{\omega_1}{\omega_2} = \frac{\Omega_1}{\Omega_2}. 
\end{equation}

\subsection{Conserved quantities}
If \(\omega_1 = \omega_2 = \omega_0\), i.e., \(\Omega_1 = \Omega_2 = 1\), then the system is clearly symmetric under transformations from \(U(2)\) \cite{2da,2db,2dc,ANISO,ANISOprev} and is therefore (maximally) superintegrable because not only are \(H_1\) and \(H_2\) conserved, we also have the conservation of angular momentum \(L = q_1 p_2 - q_2 p_1 \) and the quantity \(A = p_1p_2 + q_1q_2\); all four are not functionally independent. If \(\omega_1 \neq \omega_2\), the Hamiltonian apparently admits a smaller symmetry group, namely, \(U(1) \oplus U(1)\), which describes invariance under rotations performed individually on the \(q_1-p_1\) and \(q_2-p_2\) planes as is evident from Eq. (\ref{Hisopq}). Let us perform another set of canonical transformations as \cite{ANISO}
\begin{eqnarray}
\mathcal{X}_j(X_j,P_j)&=&\sqrt{\Omega_j}X_j^{\frac{1}{2}\left(1+\frac{1}{\Omega_j}\right)}P_j^{\frac{1}{2}\left(1-\frac{1}{\Omega_j}\right)}, \label{1x} \\
 \mathcal{P}_j(X_j,P_j)&=&\sqrt{\Omega_j}X_j^{\frac{1}{2}\left(1-\frac{1}{\Omega_j}\right)}P_j^{\frac{1}{2}\left(1+\frac{1}{\Omega_j}\right)}, \label{1p}
\end{eqnarray}
such that in these new canonical variables, the Hamiltonian reads \(H = i \omega_0 \sum_{j=1}^2 \mathcal{P}_j \mathcal{X}_j\). The conserved quantities may now be straightforwardly computed; they are simply the generators of the \(U(2)\) transformations. The conserved quantities (labeled as \(\{I_0,I_1,I_2,I_3\}\)) when expressed in the canonical variables \((q_1,q_2,p_1,p_2)\) by inverting all the canonical transformations performed on Eq. (\ref{Hisopq}) read as
\begin{eqnarray}
I_0&=&\frac{\Omega_1}{2}(p_1^2+q_1^2)+\frac{\Omega_2}{2}(p_2^2+q_2^2), \label{I0expression}\\
I_1&=&\sqrt{\Omega_1\Omega_2(p_1^2+q_1^2)(p_2^2+q_2^2)}\cos{\left[\frac{\pi}{4}\left(\frac{1}{\Omega_2}-\frac{1}{\Omega_1}\right)+\left(\frac{\Psi_2}{\Omega_2}-\frac{\Psi_1}{\Omega_1}\right)\right]},\label{addi1}\\
I_2&=&\sqrt{\Omega_1\Omega_2(p_1^2+q_1^2)(p_2^2+q_2^2)}\sin{\left[\frac{\pi}{4}\left(\frac{1}{\Omega_2}-\frac{1}{\Omega_1}\right)+\left(\frac{\Psi_2}{\Omega_2}-\frac{\Psi_1}{\Omega_1}\right)\right]},\label{addi2}\\
I_3&=&\frac{\Omega_1}{2}(p_1^2+q_1^2)-\frac{\Omega_2}{2}(p_2^2+q_2^2), \label{I3expression}
\end{eqnarray}
with $\Psi_j=-\tan^{-1}{\left(\frac{p_j}{q_j}\right)}$. 

\vspace{2mm}

It may be easily verified that the first integrals listed above Poisson-commute with the Hamiltonian which is just \(I_0\) (up to a factor of \(\omega_0\)). Further, the expressions for \(I_1\) and \(I_2\) reduce to the familiar ones for \(\Omega_1 = \Omega_2 = 1\). For example, consider \(I_2\) with \(\Omega_1 = \Omega_2 = 1\):
\begin{eqnarray}
I_2&=&\sqrt{(p_1^2+q_1^2)(p_2^2+q_2^2)}\sin (\Psi_2 - \Psi_1) \nonumber \\
&=&\sqrt{(p_1^2+q_1^2)(p_2^2+q_2^2)} \big( \sin \Psi_2 \cos \Psi_1 - \sin \Psi_1 \cos \Psi_2 \big). \label{I2isolim}
\end{eqnarray}
Since \(\tan \Psi_1 = -p_1/q_1\) and \(\tan \Psi_2 = -p_2/q_2\), one has 
\begin{equation}
\cos \Psi_1 = \frac{q_1}{\sqrt{p_1^2 + q_1^2}}, \hspace{5mm} \cos \Psi_2 = \frac{q_2}{\sqrt{p_2^2 + q_2^2}}, \hspace{5mm} \sin \Psi_1 = \frac{-p_1}{\sqrt{p_1^2 + q_1^2}}, \hspace{5mm} \sin \Psi_2 = \frac{-p_2}{\sqrt{p_2^2 + q_2^2}}.
\end{equation}
Substituting these into Eq. (\ref{I2isolim}) gives \(I_2 = q_2 p_1 - q_1 p_2\), as expected. Similarly, we get \(I_1 = p_1 p_2 + q_1 q_2\) by putting \(\Omega_1 = \Omega_2 = 1\). Notice that upon defining 
\begin{equation}\label{gammadef}
h_1 = \frac{\Omega_1}{2} (p_1^2 + q_1^2), \quad \quad h_2 = \frac{\Omega_2}{2} (p_2^2 + q_2^2), \quad \quad \Gamma = \frac{ \Omega_1 \Psi_2 - \Omega_2 \Psi_1}{\Omega_1 \Omega_2},
\end{equation} one finds that
\begin{equation}
I_0 = h_1 + h_2, \quad I_1 = 2 \sqrt{h_1 h_2} \cos (\Gamma + \Phi), \quad I_2 = 2 \sqrt{h_1 h_2} \sin (\Gamma + \Phi), \quad I_3 = h_1 - h_2,
\end{equation} where \(\Phi = \frac{\pi}{4}\left(\frac{1}{\Omega_2}-\frac{1}{\Omega_1}\right)\). Thus, one seems to have three functionally-independent first integrals, namely, \(h_1\), \(h_2\), and \(\Gamma\). Below, let us clarify some subtleties which were not pointed out by us in \cite{ANISO}. 

\subsection{Discussion}
If \(\omega_1/\omega_2\) is a rational number, i.e., the frequencies are commensurable, then the trajectories are periodic and are closed; every invariant torus is a union of periodic orbits which implies that it is foliated by invariant circles. It then makes sense to have three functionally-independent first integrals on the phase space. However, if the frequencies are incommensurable, i.e., \(\omega_1/\omega_2\) is an irrational number, then the trajectories in the phase space are only quasi-periodic and are not closed; any trajectory densely fills an invariant torus meaning that there cannot be three functionally-independent first integrals which are defined globally on a trajectory\footnote{This is consistent with the quantum-mechanical pictures of the two situations; for the two-dimensional harmonic oscillator with commensurable frequencies, one encounters the so-called `accidental degeneracy' \cite{2da} which can now be attributed to its maximal superintegrability while in the case with incommensurable frequencies, there is no accidental degeneracy.} (see for example, section (1.7) of \cite{fasso}). Notice that the transformations presented in Eqs. (\ref{1x}) and (\ref{1p}) involve raising complex variables to certain (inverse) powers pointing towards the fact that the transformations are not single-valued (see also, the older works \cite{2da,ANISOprev}). We can then have the following two situations:

\begin{enumerate}
\item If \(\omega_1/\omega_2\) is a rational number, then \(\Omega_{1,2}\) can be taken to be co-prime natural numbers (except for the isotropic case for which \(\Omega_1 = \Omega_2 = 1\)), i.e., the transformations appearing in Eqs. (\ref{1x}) and (\ref{1p}) involve fractional powers for which the roots are finite in number and may therefore map the anisotropic oscillator to the isotropic oscillator by employing branch cuts. The corresponding first integrals, i.e., \(h_1\), \(h_2\), and \(\Gamma\) can be defined globally on the trajectories; first integrals with similar expressions have appeared earlier in \cite{fasso,2da,2db}. In particular, we can write for \(\zeta_{1,2} = q_{1,2} + i p_{1,2}\), the following expression for \(\Gamma\):
\begin{eqnarray}
\Gamma &=& \frac{1}{\Omega_1\Omega_2} \bigg[\Omega_2 \tan^{-1} \bigg(\frac{p_1}{q_1}\bigg) -  \Omega_1 \tan^{-1} \bigg(\frac{p_2}{q_2}\bigg) \bigg] \nonumber \\
&=& \frac{1}{\Omega_1\Omega_2}  {\rm Im} \big[ \Omega_2 \ln \zeta_1 - \Omega_1 \ln \zeta_2 \big] \nonumber \\
&=& \frac{1}{\Omega_1\Omega_2}   {\rm Im} \bigg[ \ln \bigg(\frac{\zeta_1^{\Omega_2}}{\zeta_2^{\Omega_1}} \bigg) \bigg] .
\end{eqnarray} The quantity
\begin{equation}
\Lambda = \frac{\zeta_1^{\Omega_2}}{\zeta_2^{\Omega_1}}
\end{equation} is a first integral that is functionally independent of \(h_1\) and \(h_2\). Moreover, it can be defined globally because \(\Omega_1\) and \(\Omega_2\) are co-prime natural numbers. Thus, the anisotropic oscillator is superintegrable in the Liouville sense because one has three (independent) globally-defined first integrals to label each trajectory. 

\item If \(\omega_1/\omega_2\) is an irrational number, then one can set \(\Omega_{1} = \delta \notin \mathbb{Q}\) and \(\Omega_{2} =1\) without loss of generality. Consequently, the transformations suggested in Eqs. (\ref{1x}) and (\ref{1p}) are trivial for \(j = 2\) but involve infinitely-many roots for \(j = 1\) making the correspondence between the anisotropic oscillator and its isotropic counterpart dubious. However, \(h_1\), \(h_2\), and \(\Gamma\) Poisson-commute with the total Hamiltonian of the system indicating that they are first integrals. While \(h_1\) and \(h_2\) are defined globally, the first integral
\begin{equation}
\Gamma =   {\rm Im} \bigg[ \frac{\ln \zeta_1}{\delta}  -  \ln \zeta_2 \bigg]
\end{equation} is only defined locally due to the branch cut of the complex logarithm. 

\end{enumerate}

\section{Conserved quantities via the last-multiplier formalism}\label{sec3}
We will now describe (briefly) the formalism of the Jacobi last multiplier and its role in determining conserved quantities. For any dynamical system whose time evolution is described by a vector field \(X\), the last multiplier is a factor that satisfies \cite{2d16,JLM1,JLM7,JLM8}
\begin{equation}
\frac{d}{dt} \ln M + {\rm div}\cdot X = 0.
\end{equation}
Here, \({\rm div}\cdot X\) is the divergence of the vector field \(X\) defined in the sense that \(\pounds_X \Omega_V =( {\rm div}\cdot X) \Omega_V\), where \(\pounds_X\) is the Lie derivative with respect to the vector field \(X\) and \(\Omega_V = dx_1 \wedge dx_2 \wedge \cdots \wedge dx_m\) is the volume-form on the \(m\)-dimensional phase space in appropriate local coordinates\footnote{For Hamiltonian systems on symplectic manifolds, one should have \(m = 2n\) with \(n \in \mathbb{N}\). These local coordinates are then the Darboux coordinates in which the phase space locally looks like \(T^*\mathbb{R}^n\).} \((x_1,x_2,\cdots,x_m)\) (of course, we will be interested in orientable phase spaces). For Hamiltonian dynamics where one has the Liouville's theorem, the dynamical vector field has vanishing divergence so that \(\pounds_X \Omega_V = 0\); consequently, the last multiplier is a constant which we can set to be equal to one without loss of generality.

\vspace{2mm}

\(M\) is called the last multiplier because if for a system on an \(m\)-dimensional phase space, the last multiplier together with \(m-2\) first integrals are known, it is possible to compute the \((m-1)\)th, i.e., the `last' first integral. 
Consider a dynamical system whose dynamics is described by the vector field \(X\) with components
\begin{equation}\label{genericdynamicalsystem}
\frac{dx_j}{dt} = X_j(\{x_j\}) \quad \implies \quad X = X_j(\{x_j\}) \frac{\partial}{\partial x_j},
\end{equation} where \(j \in \{1,2,\cdots,m\}\) and \(\{x_j\}\) are some appropriate (possibly local) coordinates. If the system has \(m-2\) constants of motion, we may write them as
\begin{equation}\label{fkrel}
I_k(x_1,x_2,\cdots,x_m) = c_k, \hspace{6mm} k \in \{1,2,\cdots,m-2\}.
\end{equation} The real constants \(\{c_k\}\) are just the numerical values of the first integrals \(\{I_k\}\). One now performs a change of variables as
\begin{equation}
(x_1,x_2, \cdots, x_m) \mapsto (c_1,c_2,\cdots, c_{m-2}, \zeta_1,\zeta_2),
\end{equation} where one makes use of the relations given in Eq. (\ref{fkrel}). In doing so, one converts the system into a planar system, i.e., one now has
\begin{equation}\label{planardynamics}
\frac{d\zeta_1}{dt} = \overline{X}_1 (\zeta_1,\zeta_2, \{c_k\}), \hspace{6mm} \frac{d\zeta_2}{dt} = \overline{X}_2 (\zeta_1,\zeta_2, \{c_k\}),
\end{equation} where \(\overline{X}_1, \overline{X}_2: U \subseteq \mathbb{R}^2 \rightarrow \mathbb{R}\) are functions of the variables \((\zeta_1,\zeta_2) = (x_{m-1},x_m)\) as obtained from Eq. (\ref{genericdynamicalsystem}) via elimination of \((m-2)\) variables. Then, upon defining
\begin{equation}\label{Deltadef}
\Delta=\frac{\partial(I_1,I_2,\cdots,I_{m-2}, x_{m-1}, x_m)}{\partial(x_1,x_2,\cdots,x_{m-2}, x_{m-1}, x_m)} = \frac{\partial(I_1,I_2,\cdots,I_{m-2})}{\partial(x_1,x_2,\cdots,x_{m-2})},
\end{equation}
it follows after some manipulations and upon using the Stokes' theorem that \cite{JLM1,JLM7}
\begin{equation}\label{thetadef}
\Theta = \int \frac{\overline{M}}{\overline{\Delta}} (\overline{X}_1 d\zeta_2 - \overline{X}_2 d\zeta_1)
\end{equation} is a conserved quantity of the dynamical system given in Eq. (\ref{planardynamics}). Here, the `overline' denotes that \(\overline{M}\), \(\overline{\Delta}\), \(\overline{X}_1\), and \(\overline{X}_2\) are being considered after the \((m-2)\) variables have been eliminated.

\subsection{Revisiting the two-dimensional harmonic oscillator}
The Hamiltonian is given by Eq. (\ref{Hisopq}). Since the dynamics is of the Hamiltonian kind, we have \(M = 1\). Thus, Eq. (\ref{thetadef}) gives
\begin{eqnarray}
\Theta =\frac{\tan ^{-1}\left(\frac{q_1}{p_1}\right)}{\omega_1}-\frac{\tan ^{-1}\left(\frac{ q_2}{p_2}\right)}{\omega_2}.
\end{eqnarray}
The fact that \(\Theta\) as obtained above is indeed a first integral can be independently verified by checking that \(\{H_1+H_2,\Theta\}_{\rm P.B.}=0\), meaning \(\dot{\Theta} = 0\). The result for the isotropic oscillator may be recovered by setting \(\omega_1 = \omega_2 = \omega_0\). 

\vspace{2mm}

It turns out that \(\Theta\) has an intriguing interpretation; defining
\begin{equation}
\theta_1 = \frac{\tan ^{-1}\left(\frac{q_1}{p_1}\right)}{\omega_1}, \hspace{5mm} \theta_2 = \frac{\tan ^{-1}\left(\frac{q_2}{p_2}\right)}{\omega_2},
\end{equation} we have \(\Theta = \theta_1 - \theta_2\), where \(\theta_1\) is a function of \((q_1,p_1)\) and \(\theta_2\) is a function of \((q_2,p_2)\). One can verify that
\begin{equation}
\{\theta_1 , H_1\}_{\rm P.B.} = \{\theta_2,H_2\}_{\rm P.B.} = 1,
\end{equation} i.e., one can perform the canonical transformations \((q_1,p_1) \mapsto (\theta_1,H_1)\) and \((q_2,p_2) \mapsto (\theta_2,H_2)\) in which \(H_1\) and \(H_2\) are conserved. But these are precisely the action-angle variables meaning that \(\Theta = \theta_1 - \theta_2\) is just the difference between the two angle variables \cite{2d16}.

\section{Physical examples}\label{sec4}
In this section, we shall apply the formalism of the last multiplier to compute additional first integrals of some two-dimensional systems. 

\subsection{Classical Landau problem with a hyperbolic mode}\label{magsec}
Consider the two-dimensional dynamics of a charged particle in an electromagnetic field. The Hamiltonian takes the generic form which goes as
\begin{equation}
H = \frac{(\mathbf{p} - e\mathbf{A})^2}{2m} + e\phi,
\end{equation} where \(\mathbf{p} = (p_x,p_y)\) is the canonical momentum (vector), \(\mathbf{A} = (A_x,A_y)\) is the vector potential, \(\phi\) is the scalar potential, and \(e\) is the electric charge; \(A_x\), \(A_y\), and \(\phi\) are functions of \((x,y)\). The magnetic field points in the direction perpendicular to the plane of motion and is given by \(B = \partial_x A_y - \partial_y A_x\). Let us choose the Landau gauge in which \(A_x = 0\) and \(A_y = Bx\), where \(B\) is a real constant; the Hamiltonian subsequently reads
\begin{equation}\label{Hlandau1}
H = \frac{p_x^2}{2m} + \frac{(p_y - m \omega_c x)^2}{2m} + e\lambda xy,
\end{equation} where \(\omega_c = eB/m\) and we also chose \(\phi = \lambda xy\) with \(\lambda > 0\) being a constant, following \cite{landau}. This particular problem has been found to be relevant in the context of the Riemann zeroes \cite{landau}. We shall demonstrate the existence of three constants of motion in the limit of strong magnetic field in which the Hamiltonian may be separated as
\begin{eqnarray}\label{Hlandau2}
    H=\frac{\omega_c}{2} \left(p^2+q^2\right)+|\omega_h| P Q,
\end{eqnarray} where \(\omega_h = i \lambda /B\) and \(\omega_c >> |\omega_h|\); \(\omega_c\) and \(|\omega_h|\) are the frequencies associated with the cyclotronic and hyperbolic modes, respectively. In arriving from Eq. (\ref{Hlandau1}) to (\ref{Hlandau2}), one uses the following canonical transformations:
\begin{equation}
q = x + p_y, \quad p = p _x, \quad Q = -p_y, \quad P = y + p_x,
\end{equation} in conjunction with the limit of strong magnetic field. With the separable Hamiltonian [Eq. (\ref{Hlandau2})] in hand, one can straightforwardly identify two conserved quantities, namely, 
\begin{equation}
I_1 = \frac{\omega _c}{2} \left(p^2+q^2\right), \quad\quad I_2 = |\omega_h| P
Q.
\end{equation}
Using Eq. (\ref{thetadef}), we can find the third conserved quantity which turns out to be
\begin{eqnarray}
\Theta=\frac{2}{\omega _c}\tan ^{-1}\left(\frac{q}{p}\right)+\frac{2}{|\omega_h|} \tanh ^{-1}\left(\frac{P-i Q}{P+i Q}\right),
\end{eqnarray} which indeed `commutes' with the Hamiltonian with respect to the Poisson bracket. Reverting back to the original variables \((x,y,p_x,p_y)\), we have the following expression:
\begin{eqnarray}
\Theta=\frac{2}{\omega _c}\tan ^{-1}\left(\frac{x+p_y}{p_x}\right)+\frac{2}{|\omega_h|} \tanh ^{-1} (e^{2i \theta}), \quad\quad \theta = \tan^{-1} \bigg( \frac{p_y}{y + p_x} \bigg) .
\end{eqnarray}

\subsection{Two-dimensional Kepler problem}\label{kepsec}
Let us consider the two-dimensional version of the Kepler problem. In the plane-polar coordinates \((r,\psi)\), the Hamiltonian reads as (we will take the particle to have unit mass)
\begin{equation}
H=\frac{p_r^2}{2}+\frac{p_{\psi }^2}{2 r^2}-\frac{k}{r},
\end{equation} where \(k > 0\) is a constant. Because \(\psi\) is a cyclic coordinate, one finds that \(p_\psi\) is conserved thus giving us two conserved quantities to begin with. Then, Eq. (\ref{thetadef}) gives
\begin{eqnarray}\label{thetaJLMkeppler}
\Theta = \cot ^{-1}\left(\frac{ r p_r p_{\psi }}{-k  r+p_{\psi
}^2}\right)+\psi.
\end{eqnarray}
Physically, the conserved quantity determined above is related to the Laplace-Runge-Lenz vector. To uncover this connection, let us begin by noting that the two components of the Laplace-Runge-Lenz vector on the plane read as
\begin{equation}
\mathcal{A}_x = \dot{y} (x \dot{y} - y \dot{x}) - \frac{kx}{r}, \quad \quad \mathcal{A}_y = -\dot{x} (x \dot{y} - y \dot{x}) - \frac{ky}{r},
\end{equation} with \(r = \sqrt{x^2 + y^2}\). In plane-polar coordinates with \(x = r\cos \psi\) and \(y = r\sin \psi\), we have
\begin{equation}
\mathcal{A}_x = p_r p_\psi \sin \psi + \bigg(\frac{p_\psi^2}{r} - k \bigg)\cos \psi, \quad \quad \mathcal{A}_y = -p_r p_\psi \cos \psi + \bigg(\frac{p_\psi^2}{r} - k \bigg)\sin \psi. 
\end{equation}
Thus, we can define the ratio between the two components of the Laplace-Runge-Lenz vector as
\begin{eqnarray}
\frac{\mathcal{A}_x}{\mathcal{A}_y} &=&  \frac{r p_r p_\psi \sin \psi + (p_\psi^2 - kr) \cos \psi}{-r p_r p_\psi \cos \psi + (p_\psi^2 - kr) \sin \psi} \nonumber \\
&=& \frac{(\alpha/\beta) + \cot \psi}{1 - (\alpha/\beta) \cot \psi},
\end{eqnarray} where we have defined \(\alpha = r p_r p_\psi\) and \(\beta = p_\psi^2 - kr\). Eq. (\ref{thetaJLMkeppler}) implies that \(\alpha/\beta = \cot (\Theta - \psi)\), giving
\begin{eqnarray}
\frac{\mathcal{A}_x}{\mathcal{A}_y} &=&  \frac{\cot (\Theta - \psi) + \cot \psi}{1 - \cot (\Theta - \psi) \cot \psi} \nonumber \\
&=& - \tan \Theta. 
\end{eqnarray} 
Thus, the conserved quantity \(\Theta\) given in Eq. (\ref{thetaJLMkeppler}) is related to the ratio of the two components of the Laplace-Runge-Lenz vector. 

\subsection{Linear curl forces}\label{curlsec}
As our final example, we will consider a mechanical problem with linear curl forces. A curl force is one that cannot be derived as a gradient of a scalar potential, i.e., it has a non-trivial curl. We refer the reader to the works of Berry and Shukla (see for example, \cite{berry,berry1}) on curl forces and consider from \cite{berry} a simplified version of a specific example presented therein. On the plane and in Cartesian coordinates, the components of the force read as
\begin{equation}
F_x = - \mu y, \quad\quad F_y = \mu x, \quad \quad \mu > 0. 
\end{equation} The curl (vorticity) reads as \(\Omega = \partial_x F_y - \partial_y F_x = 2\mu\) and is non-vanishing. The dynamics is described by a vector field that goes as (we will take the particle to have unit mass)
\begin{equation}
X = F_x \frac{\partial}{\partial v_x} + F_y \frac{\partial}{\partial v_y} + v_x \frac{\partial}{\partial x} + v_y \frac{\partial}{\partial y},
\end{equation} such that given any function \(f = f (x,y,v_x,v_y)\), one has \(X(f) = \dot{f}\). It is straightforward to verify in the Cartesian coordinates as introduced above that \({\rm div}\cdot X = 0\), meaning that the last multiplier can be set to one. In \cite{berry}, the following two conserved quantities were introduced: 
\begin{equation}
I_1 = \frac{1}{2} (v_x^2 - v_y^2) + \mu xy, \quad \quad I_2 = v_x v_y - \frac{\mu}{2} (x^2 - y^2),
\end{equation} and it may be verified by explicit calculation that \(X(I_1) = X(I_2) = 0\). Equipped with these, we may proceed towards deriving the third constant of motion. Let us define 
\begin{equation}
Q_1 = I_1 + i I_2 = \frac{1}{2} (w^2 - i\mu z^2),
\end{equation} where \(z = x + iy\) and \(w = \dot{z} = v_x + i v_y\). Similarly, we get another functionally-independent conserved quantity to be \begin{equation}
Q_2 = Q_1^* = \frac{1}{2} ({w^*}^2 + i\mu {z^*}^2).
\end{equation} 
Thus, we have
\begin{equation}
\overline{\Delta} = {\rm det} \begin{pmatrix}
\frac{\partial I_1}{\partial w}&\frac{\partial I_2}{\partial w}\\
\frac{\partial I_1}{\partial {w^*}}&\frac{\partial I_2}{\partial {w^*}}
\end{pmatrix} = {\rm det}
\begin{pmatrix}
w&0\\
0&w^*
\end{pmatrix} ,
\end{equation} giving \(\overline{\Delta} = w w^* \). We therefore get the third conserved quantity to be 
\begin{eqnarray}
\Theta &=& \int \frac{1}{w w^*} (\dot{z}^* dz - \dot{z} dz^*) \nonumber \\
&=& \int \frac{1}{w w^*} (w^* dz - w dz^*) \nonumber \\
&=& \int \bigg(\frac{dz}{w} - \frac{dz^*}{w^*}\bigg) \nonumber \\
&=& \tanh^{-1} \bigg[ \frac{\sqrt{\mu} (-1)^{3/4} z^*}{w^*} \bigg] - i \tanh^{-1} \bigg[ \frac{\sqrt{\mu} (-1)^{1/4} z}{w} \bigg]. 
\end{eqnarray}

\section{Closing remarks}\label{dsec}
This paper has expanded upon some previous developments by the present authors \cite{ANISO,2d16}. On one hand, we have discussed the conserved quantities of the two-dimensional harmonic oscillator. On the other hand, we have described the method of the last multiplier for computing additional first integrals of some two-dimensional Hamiltonian systems (on four-dimensional phase spaces). It must be pointed out that the first integral that is obtained using the formalism of the last multiplier is often only defined locally.   \\

\section*{Acknowledgements} We thank the anonymous referees for their valuable remarks. A. G. is grateful to Ond\v{r}ej Kub\r{u} and Libor \v{S}nobl for some important remarks and also thanks Miloslav Znojil for hospitality during his visit to the Czech Technical University in Prague. The work of A. G. is supported by the Ministry of Education (MoE), Government of India in the form of a Prime Minister's Research Fellowship (ID: 1200454). A. S. would like to acknowledge the financial support from IIT Bhubaneswar in the form of an Institute Research Fellowship. B. B. thanks Brainware University for infrastructural support. We are thankful to the organizers (in particular, to \v{C}estm\'ir Burd\'ik) of the {\it XXVIII. International Conference on Integrable Systems and Quantum Symmetries} at the Czech Technical University in Prague for the opportunity to present this work. A. G. is grateful to the Department of Physics, IIT Jodhpur for hospitality during the final stages of preparing the manuscript.



\begin{thebibliography}{99}



\bibitem{fasso} Fass\`o F 1999 {\it Notes on Finite Dimensional Integrable Hamiltonian Systems} (Padua: Universit\`a di Padova)

\bibitem{evans} Evans N W 1990 
{\it Phys. Rev. A} \textbf{41} 5666


\bibitem{marquette} Marquette I 2012 
{\it J. Math. Phys.} \textbf{53} 012901

\bibitem{ANISO}
Sinha A, Ghosh A and Bagchi B 2023
{\it Phys. Scr.} \textbf{98} 095253


\bibitem{2d16}
Sinha A and Ghosh A 2024
	\textit{Pramana} \textbf{98} 101



\bibitem{2da} Dulock V A and McIntosh H V 1965 
{\it Am. J. Phys.} \textbf{33} 109

\bibitem{2db} Maiella G and Vitale B 1967 
{\it Nuovo Cimento A} \textbf{47} 330

\bibitem{2dc} Maiella G 1967 
{\it Nuovo Cimento A} \textbf{52} 1004

\bibitem{ANISOprev}
Amiet J P and Weigert S 2002
{\it J. Math. Phys.} \textbf{43} 4110







\bibitem{JLM1}
Whittaker E T 1988
{\it A Treatise on the Analytical Dynamics of Particles and Rigid Bodies}
(Cambridge: Cambridge University Press)
 
 \bibitem{JLM7}
 Guha P and Ghose Choudhury A 2013
 {\it Rev. Math. Phys.} \textbf{25} 1330009

 \bibitem{JLM8}
Cari\~nena J F and Fern\'andez-N\'u\~nez J 2021
 {\it Symmetry} \textbf{13} 1413
 
 
 \bibitem{landau}
Sierra G and Townsend P K 2008
{\it Phys. Rev. Lett.} \textbf{101} 110201

 \bibitem{berry}
Berry M V and Shukla P 2015
{\it Proc. R. Soc. A} \textbf{471} 20150002

  
 \bibitem{berry1}
Berry M V and Shukla P 2016 
{\it New J. Phys.} \textbf{18} 063018
 

	






\end{thebibliography}
\end{document}